\documentclass[onecolumn]{aa}

\usepackage{graphicx}
\usepackage{txfonts}

\begin{document}

\title{
Numerical simulation of surface brightness of astrophysical jets}

\author{
Carl~L. Gardner \and
Jeremiah~R. Jones \and
Perry~B. Vargas}

\institute{School of Mathematical \& Statistical Sciences,
Arizona State University, Tempe AZ 85287\\
\email{carl.gardner@asu.edu, jrjones8@asu.edu, perry.vargas@asu.edu}}

\abstract{We outline a general procedure for simulating the surface
  brightness of astrophysical jets (and other astronomical objects) by
  post-processing gas dynamical simulations of densities and
  temperatures using spectral line emission data from the
  astrophysical spectral synthesis package {\em Cloudy}.  Then we
  validate the procedure by comparing the simulated surface brightness
  of the HH~30 astrophysical jet in the forbidden [O~I], [N~II], and
  [S~II] doublets with {\em Hubble Space Telescope}\/ observations of
  Hartigan and Morse and multiple-ion magnetohydrodynamic simulations
  of Tesileanu et al.  The general trend of our simulated surface
  brightness in each doublet using the gas dynamical/{\em Cloudy}\/
  approach is in excellent agreement with the observational data.}


\maketitle

\section{Introduction}

To understand the physics of astrophysical jets, especially near their
sources, requires -- in addition to densities, temperatures, and
velocities -- knowledge of surface brightness of the jets in different
spectral lines.  Simulated surface brightness maps in various lines
may be constructed by applying gas dynamics or magnetohydrodynamics
(MHD) with multiple ionic species and calculating the spectral
emission from ionization effects in radiative cooling ``on the fly,''
as in the pioneering MHD computations of Tesileanu et al.\ (2014); or
by applying gas dynamics (or MHD) with radiative cooling and then
post-processing densities and temperatures using an astrophysical
spectral synthesis package like {\em Cloudy}\/ (version 13.03, last
described by Ferland et al.\ 2013) to model the spectral emission (see
Stute et al.\ (2010) for a very different application of this
procedure, using a different set of software tools).  In Gardner et
al.\ 2016, surface brightness maps from our gas dynamics/{\em
  Cloudy}\/ simulations of the SVS 13 micro-jet -- traced by the
emission of the shock excited 1.644 $\mu$m [Fe~II] line -- and bow
shock bubble -- traced in the lower excitation 2.122 $\mu$m H$_2$
line -- projected onto the plane of the sky are shown to be in good
qualitative agreement with (non-quantitative) observational images in
these lines.  This investigation will quantitatively compare simulated
surface brightness results for the HH~30 jet of the more fundamental
approach of Tesileanu et al.\ (2014) with our more phenomenological
gas dynamics/{\em Cloudy}\/ approach and with {\em Hubble Space
  Telescope}\/ ({\em HST}) observations of Hartigan and Morse (2007).

The Herbig-Haro object HH~30 consists of a pair of jets from a young
star in the Taurus star formation region at a distance of 140 pc
(Kenyon et al.\ 1994).  {\em HST} images display a nearly edge-on,
flared, reflection nebula on both sides of an opaque circumstellar
disk, with collimated jets emitted perpendicular to the disk on both
sides and nearly in the plane of the sky (the declination angle is
approximately 80$^\circ$ with respect to the line of sight (Hartigan
and Morse 2007)).  The HH~30 jets extend out to about 0.25 pc from the
stellar source in each direction (Lopez et al.\ 1996), but here we are
concerned with the first 0.003 pc of the brighter (northeastern)
blueshifted jet near its launch site.  Since the nearly edge-on disk
effectively blocks the stellar light, the jet can be studied as it
emerges from the accretion disk.

We validate our gas dynamics/{\em Cloudy}\/ procedure by comparing the
simulated surface brightness of the HH~30 astrophysical jet in the
forbidden [O~I], [N~II], and [S~II] doublets with {\em HST}\/
observations of Hartigan and Morse (2007) and simulations of Tesileanu
et al.\ (2014).  On a logarithmic scale, our surface brightness
simulations along the center of the jet are always within $\pm 7.5$\%
of the observational data for all three doublets, and fit the
observational data more tightly than the simulations of Tesileanu et
al.\ (2014).  (However, it is important to note that the aim of
Tesileanu et al.\ (2014) was quite different: these authors simulated
a generic MHD jet with radiative cooling and spectral emission from
multiple ionic species, and then compared their generic simulation
with the observational data from HH~30 and RW Aurigae.)  More
importantly, the general trend of our simulated surface brightness in
each doublet closely follows the observational patterns, and we can
simulate surface brightness all the way up to the circumstellar disk.

\section{Numerical methods and radiative cooling}

To simulate astrophysical jets, we apply the WENO (weighted
essentially non-oscillatory) method (Shu 1999) -- a modern high-order
upwind method -- to the equations of gas dynamics with atomic and
molecular radiative cooling (see Ha et al.\ 2005, Gardner \& Dwyer
2009, Gardner et al. 2016).  The equations of gas dynamics with
radiative cooling can be written as hyperbolic conservation laws for
mass, momentum, and energy:
\begin{equation}
	\frac{\partial \rho}{\partial t} + 
	\frac{\partial}{\partial x_i} (\rho u_i) = 0
\label{n}
\end{equation}
\begin{equation}
	\frac{\partial}{\partial t} (\rho u_j) + \frac{\partial}{\partial x_i}
	(\rho u_i u_j) + \frac{\partial P}{\partial x_j} = 0
\label{p}
\end{equation}
\begin{equation}
	\frac{\partial E}{\partial t} + \frac{\partial}{\partial x_i}
	\left(u_i (E + P)\right) = - C(n, T) ,
\label{E}
\end{equation}
where $\rho = m n$ is the density of the gas, $m$ is the average mass
of the gas atoms or molecules, $n$ is the number density, $u_i$ is the
velocity, $\rho u_i$ is the momentum density, $P = n k_B T$ is the
pressure, $T$ is the temperature, and $E$ is the energy density of the
gas.  Indices $i,~ j$ equal 1, 2, 3, and repeated indices are summed
over.  The pressure is computed from the other state variables by the
equation of state:
\begin{equation}
	P = (\gamma - 1)\left(E - \frac{1}{2} \rho u^2\right) ,
\label{EOS}
\end{equation}
where $\gamma$ is the polytropic gas constant.

We use a ``one fluid'' approximation and assume that the gas is
predominantly H above 8000~K, with the standard admixture of the most
abundant elements in the interstellar medium (ISM); while below
8000~K, we assume the gas is predominantly H$_2$, with $n(H)/n(H_2)
\approx 0.01$.  We make the further approximation that $\gamma$ equals
$\frac{5}{3}$ (the value for a monatomic gas) for simplicity.

\begin{figure}
\begin{center}
\includegraphics[scale=0.35]{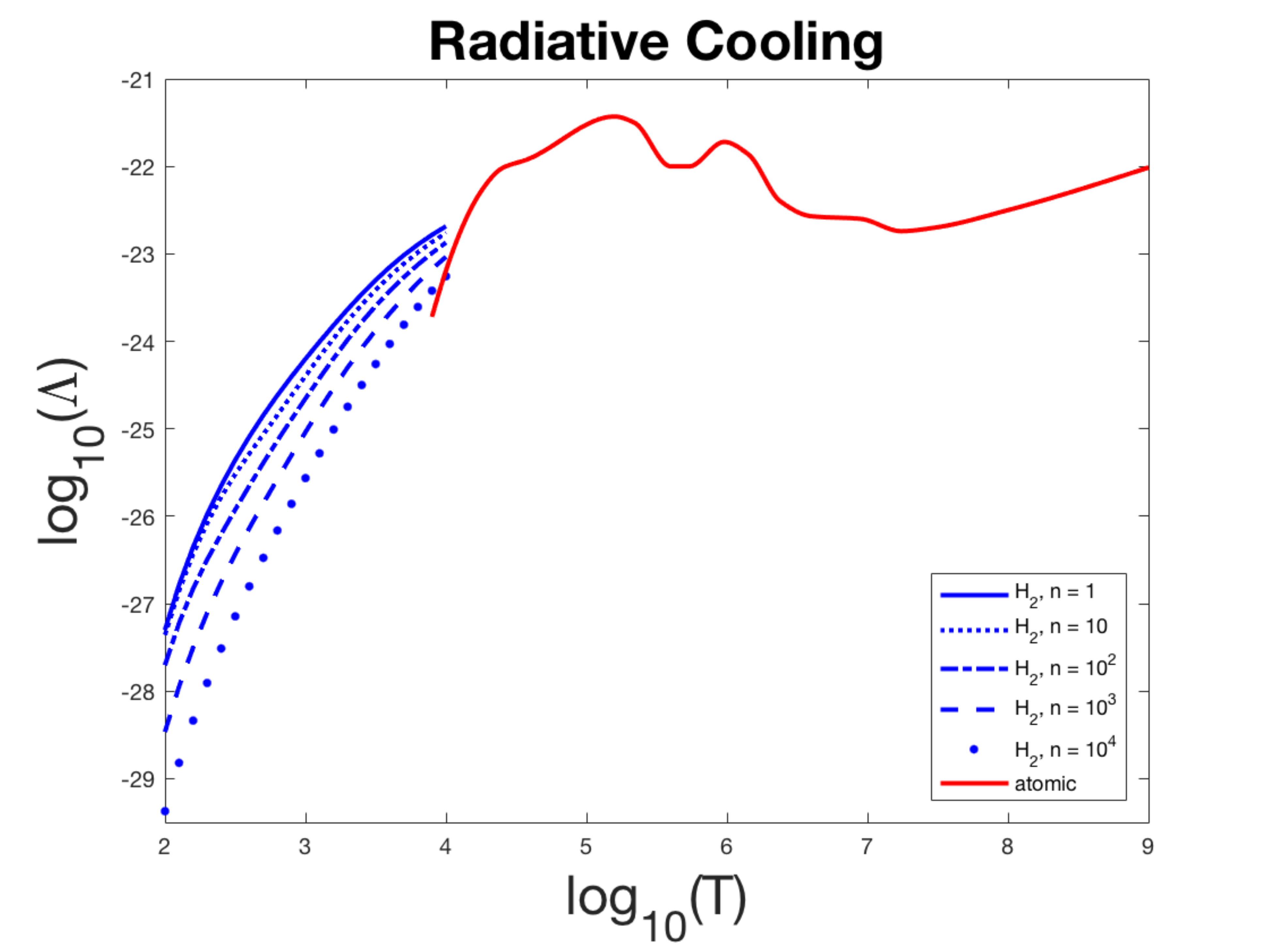}
\caption{Atomic $\Lambda(T)$ and molecular $\Lambda(n,T) = W(n,T)/n$
  cooling functions: $\log_{10}(\Lambda)$ with $\Lambda$ in erg cm$^3$
  s$^{-1}$, $n$ in H atoms/cm$^3$ ($n$ = 100 corresponds to 50 H$_2$
  molecules/cm$^3$), and $T$ in K\@.}
\label{fig:Mol-Atom-Cooling}
\end{center}
\end{figure}

Radiative cooling (see Fig.~\ref{fig:Mol-Atom-Cooling}) of the gas
is incorporated through the right-hand side $-C(n, T)$ of the energy
conservation equation~(\ref{E}), where
\begin{equation}
        C(n, T) = \left\{ \begin{array}{ll}
        n^2 \Lambda(T) & T \ge 8000~{\rm K,~ for~atomic~cooling~only}\\
        n W(n, T) & T < 8000~{\rm K,~ for~H}_2~{\rm cooling~only}\\
	\end{array} \right.
\label{cooling}
\end{equation}
with the model for $\Lambda(T)$ taken from Fig.~8 of Schmutzler \&
Tscharnuter (1993) for atomic cooling, and the model for $W(n, T)$
from Fig.~4 of Le Bourlot et al.\ (1999) for H$_2$ molecular cooling.
The atomic cooling model includes the relevant emission lines of the
ten most abundant elements (H, He, C, N, O, Ne, Mg, Si, S, Fe) in the
ISM, as well as relevant continuum processes.  Both atomic and
molecular cooling are actually operative between 8000~K $\le T \le$
10,000~K, but atomic cooling is dominant in this range.

We use a positivity preserving (Hu et al.\ 2013) version of the
third-order WENO method (Shu 1999, p.~439) for the gas dynamical
simulations (Gardner et al.\ 2017).  ENO and WENO schemes are
high-order, upwind, finite-difference schemes devised for nonlinear
hyperbolic conservation laws with piecewise smooth solutions
containing sharp discontinuities like shock waves and contacts.  Most
numerical methods for gas dynamics can produce negative densities and
pressures, breaking down in extreme circumstances involving very
strong shock waves, shock waves impacting molecular clouds, strong
vortex rollup, etc.  By limiting the numerical flux, positivity
preserving methods guarantee that the gas density and pressure always
remain positive.

Below we calculate emission maps from the 2D cylindrically symmetric
simulations of the HH~30 jet, at a distance of $R$ = 140 pc.  To
calculate the [O~I] (6300 + 6363~\AA), [N~II] (6584 + 6548~\AA), and
[S~II] (6716 + 6731~\AA) doublet emission, we post-processed the
computed solutions using emissivities $\epsilon(n,T)$ extracted and
tabulated from {\em Cloudy}.  In practice, we calculate (see
Table~\ref{table:code}) a table of values for
$\log_{10}(\epsilon_{line})$ on a grid of $(\log_{10}(n),
\log_{10}(T))$ values relevant in the simulations to each line, and
then use bilinear interpolation in $\log_{10}(n)$ and $\log_{10}(T)$
to compute $\log_{10}(\epsilon_{line})$.  For the simulations
presented here, we calculated the emissivities for $2 \le \log_{10}(n)
\le 8$ with $n$ in H atoms/cm$^3$ with a spacing of 0.1 in
$\log_{10}(n)$; and for $3.8 \le \log_{10}(T) \le 6$ with $T$ in K with
a spacing of 0.05 in $\log_{10}(T)$.

\begin{table}[htb]
\caption{{\em Cloudy}\/ code for calculating emissivities $\epsilon(n,T)$.}
\label{table:code}
\center{
\begin{tabular}{|l|} \hline
save grid \texttt{"}.grd\texttt{"} \\ 
c create ISM radiation field background \\ 
table ISM \\ 
abundances ISM \\ 
c vary Hydrogen density in powers of 10 \\ 
hden 2 vary \\ 
grid 2 8 0.1 \\ 
c vary temperature in powers of 10 \\ 
constant temperature 3 vary \\ 
grid 3.8 6 0.05 \\ 
c stop at zone 1 for speed \\ 
stop zone 1 \\ 
c save emissivity \\ 
save lines emissivity \texttt{"}.ems\texttt{"} \\ 
O 1 6300A \\ 
O 1 6363A \\
N 2 6584A \\ 
N 2 6548A \\ 
c SII 6716A + 6731A: \\
S 2 6720A \\ 
end of lines \\
\hline
\end{tabular}
}
\end{table}

Surface brightness $S_{line}$ in any emission line is calculated by
integrating along the line of sight through the jet and its
surroundings
\begin{equation}
        S_{line} = \frac{\int \epsilon_{line}(n, T) dl}{4 \pi R^2}
\end{equation}
where $R$ is the distance to the jet, and then converting to erg
cm$^{-2}$ arcsec$^{-2}$ s$^{-1}$.

\section{Results of numerical simulations}
\label{sec:results}

Parallelized cylindrically symmetric simulations were performed on a
$750 \Delta z \times 100 \Delta r$ grid, spanning $1.5 \times 10^{11}$
km by $0.4 \times 10^{11}$ km (using the cylindrical symmetry).  The
jet was emitted through a disk-shaped inflow region in the $r z$ plane
centered on the $z$-axis with a diameter of $10^{9}$ km, and
propagated along the $z$-axis with an initial velocity of 300 km
s$^{-1}$.  The simulation parameters at $t = 0$ for the jet and
ambient gas are given in Table~\ref{table:params}.  The jet is
propagating into previous outflows, so although the far-field ambient
is lighter than the jet, the immediate ambient near the stellar source
is likely to be nearly as dense as the jet itself.  Further, the
tapering morphology of the {\em HST}\/ observations of the jet imply
that the jet density is near the ambient density: if the jet is much
heavier than the ambient, the jet would create a strong bow shock and
would itself be wider; if the jet is much lighter than the ambient,
the jet would also create a strong bow shock, plus strong
Kelvin-Helmholtz rollup and entrainment of the ambient gas.  Note that
the jet and ambient gas are not pressure matched.

\begin{table}[htb]
\caption{Initial parameters for the jet and ambient gas.}
\label{table:params}
\center{
\begin{tabular}{l l} \hline \hline 
Jet & Ambient Gas \\ \hline 
$n_j$ = $10^5$ H/cm$^3$ & $n_a$ = $2.5 \times 10^4 $ H$_2$/cm$^3$ \\ 
$u_j$ = 300 km s$^{-1}$ & $u_a$ = 0 \\ 
$T_j$ = $10^4$ K & $T_a$ = $10^3$ K \\ 
\hline
\end{tabular}
}
\end{table}

The jet is periodically pulsed in order to approximate the
observational density and temperature data in Hartigan and Morse
(2007), with the pulse on for 6.5 yr and then off for 1.5 yr over the
simulated time of 35 yr, giving four full pulses plus a final partial
pulse.  We then spatially match the partial pulse plus the next three
full pulses of the jet (projected with a declination angle of
80$^\circ$ with respect to the line of sight) out to $0.94 \times
10^{11}$ km = 4.5 arcsec with the {\em HST}\/ observations.  Our
approach was to vary the jet and ambient parameters and the pulses to
obtain good agreement with the {\em HST}\/ observations of density and
temperature, and only then to calculate the surface brightness of the
jet in the three forbidden doublets.

\begin{figure*}
\begin{center}
\includegraphics[scale=0.45]{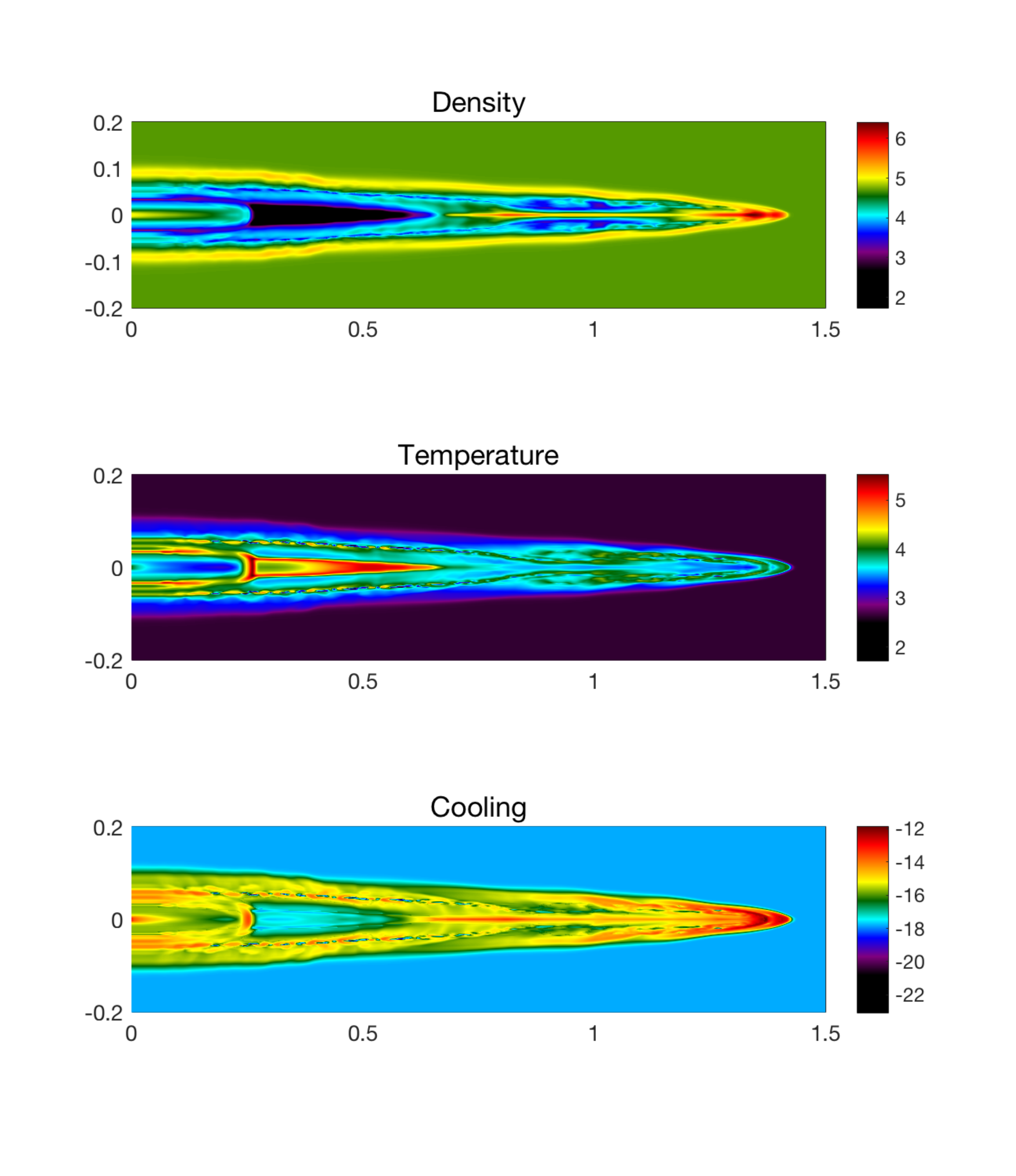}
\vspace*{-1in}
\caption{Logarithm of density $\log_{10}(n)$ with $n$ in H
  atoms/cm$^3$ (top panel), logarithm of temperature $\log_{10}(T)$
  with $T$ in K (middle panel), and logarithm of total radiative
  cooling $\log_{10}(C)$ with $C$ in erg cm$^{-3}$ s$^{-1}$ (bottom
  panel) at $t$ = 35 yr.  Lengths along the boundaries are in units of
  $10^{11}$ km.  The {\em HST}\/ observations extend out to $0.94
  \times 10^{11}$ km = 4.5 arcsec.}
\label{fig:jet}
\end{center}
\end{figure*}

\begin{figure*}
\begin{center}
\includegraphics[scale=0.45]{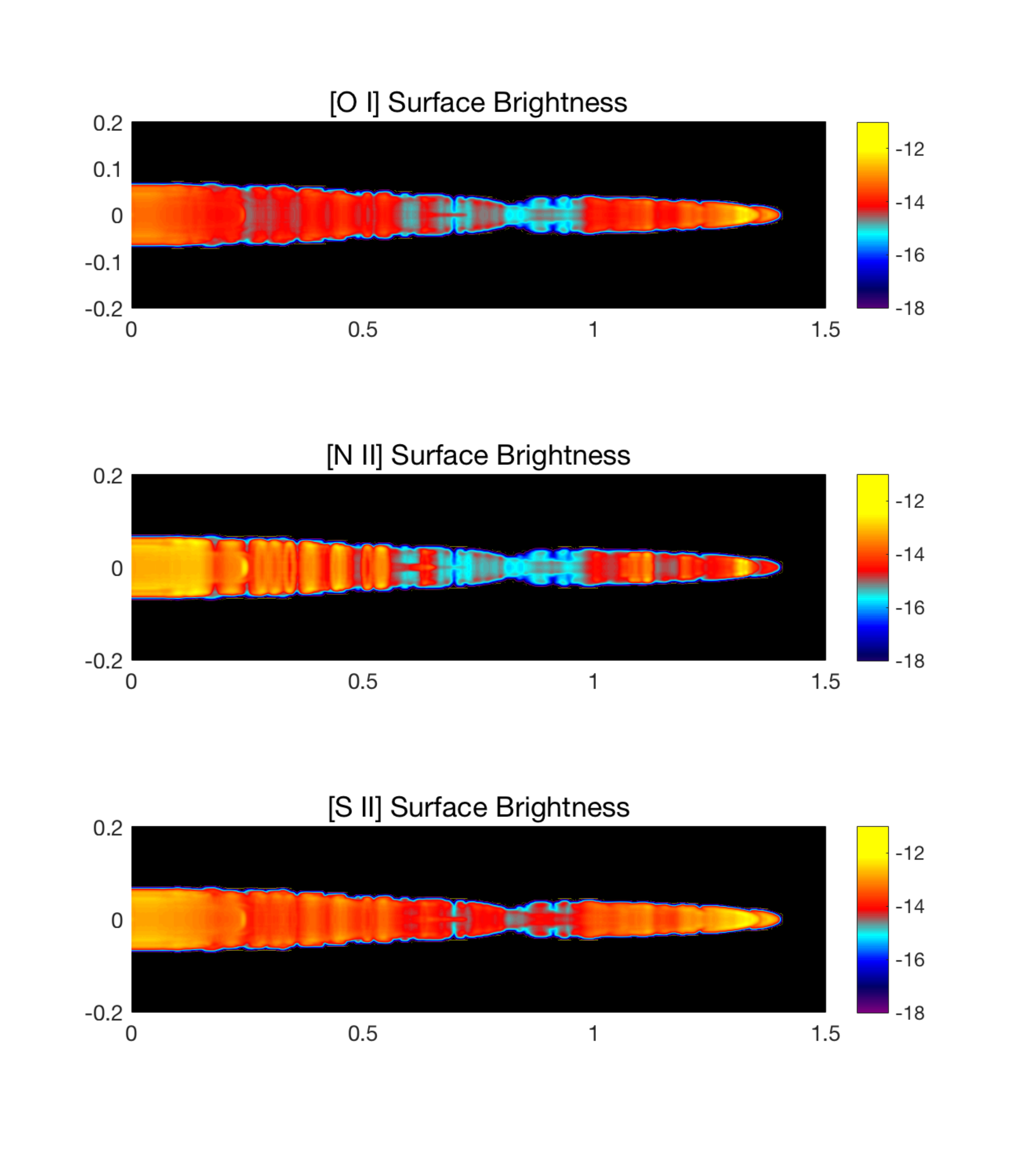}
\vspace*{-1in}
\caption{Simulated surface brightness $S$ in the [O~I] (6300 +
  6363~\AA) doublet (top panel), [N~II] (6584 + 6548~\AA) doublet
  (middle panel), and [S~II] (6716 + 6731~\AA) doublet (bottom panel)
  at 35 yr, projected 80$^\circ$ with respect to the line of sight:
  $\log_{10}(S)$ with $S$ in erg cm$^{-2}$ arcsec$^{-2}$ s$^{-1}$.
  Lengths along the boundaries are in $10^{11}$ km.  The {\em HST}\/
  observations extend out to $0.94 \times 10^{11}$ km = 4.5 arcsec.}
\label{fig:SB3plots}
\end{center}
\end{figure*}

In the simulation Figs.~\ref{fig:jet} and~\ref{fig:SB3plots}, the jet
is surrounded by a thin bow shock plus thin bow-shocked cocoon.  The
jet is propagating initially at Mach 25 with respect to the sound
speed in the jet/ambient gas.  However, the jet tip actually
propagates at an average velocity of approximately 110 km s$^{-1}$, at
Mach 8 with respect to the average sound speed in the ambient gas,
since the jet is slowed down as it impacts the ambient environment.
The average velocity of gas within the jet is approximately 200 km
s$^{-1}$, in excellent agreement with the {\em HST}\/ observations
(Bacciotti, et al.\ 1999).

The HH~30 jet is one of the densest jets observed (Bacciotti et
al.\ 1999).  Observationally the jet density starts at $10^5$
H/cm$^3$, decreases to $5 \times 10^4$ H/cm$^3$ within first arcsec,
and then gradually falls to $10^4$ H/cm$^3$ at a few arcsec; the jet
temperature starts at $2 \times 10^4$~ K, decreases to $10^4$~K within
first arcsec, and then gradually decays to 6000--7000~K at a few
arcsec.  These data are well matched by our simulated densities and
temperatures in Fig.~\ref{fig:jet} out to the limit of the {\em HST}\/
observations at $0.94 \times 10^{11}$ km = 4.5 arcsec.

Figure~\ref{fig:SB3plots} presents our simulated surface brightness of
the jet in the three forbidden doublets, projected onto the plane of
the sky.  The surface brightness has been smoothed with a Gaussian
point-spread function with a width of 0.1 arcsec.  Qualitatively the
simulations are similar to the observations of Hartigan and Morse
(2007), but the real test is to compare surface brightness in each
doublet along the center of the jet.

\section{Discussion}
\label{sec:discuss}

As is evident in Figs.~\ref{fig:jet} and~\ref{fig:SB3plots}, the jet
pulses create a series of shocked knots followed by rarefactions
within the jet.  In Fig.~\ref{fig:jet}, the panels indicate shocked
knots at $z \approx 0.25$, 0.8, 1.0, and $1.4 \times 10^{11}$ km, with
an incipient knot beginning at $z = 0$.  The incipient knot plus the
next three knots correspond roughly to the N, A, B, C knots
illustrated in Hartigan and Morse (2007), while the final knot is
outside the spatial range of their observations.  In our simulations,
the jet creates a terminal Mach disk near the jet tip, which reduces
the velocity of the jet flow to the flow velocity of the contact
discontinuity at the leading edge of the jet.  With radiative cooling,
the jet exhibits a higher density contrast near its tip (when the
shocked, heated gas cools radiatively, it compresses), a narrower bow
shock, and lower overall temperatures.  There is a strong rarefaction
between the knots for $z \approx 0.25$--$0.7 \times 10^{11}$ km.  The
knots at $z \approx 0.25$, $0.8$, and $1.1 \times 10^{11}$ km are
undergoing especially strong radiative cooling.  The temperature is
highest around the knot at $z \approx 0.25$ and for the strong
rarefaction between $z \approx 0.25$--$0.7 \times 10^{11}$ km.  In
Fig.~\ref{fig:SB3plots}, there is strong cooling in all three doublets
for $0 \le z \le 0.6 \times 10^{11}$ km and for $1.0 \times 10^{11}
\le z \le 1.4 \times 10^{11}$ km.  The [O~I] doublet emission is low
for $0.8 \times 10^{11} \le z \le 1.0 \times 10^{11}$ km, while the
[N~II] doublet emission is low for $0.7 \times 10^{11} \le z \le 1.0
\times 10^{11}$ km.

\begin{figure}
\begin{center}
\includegraphics[scale=0.35]{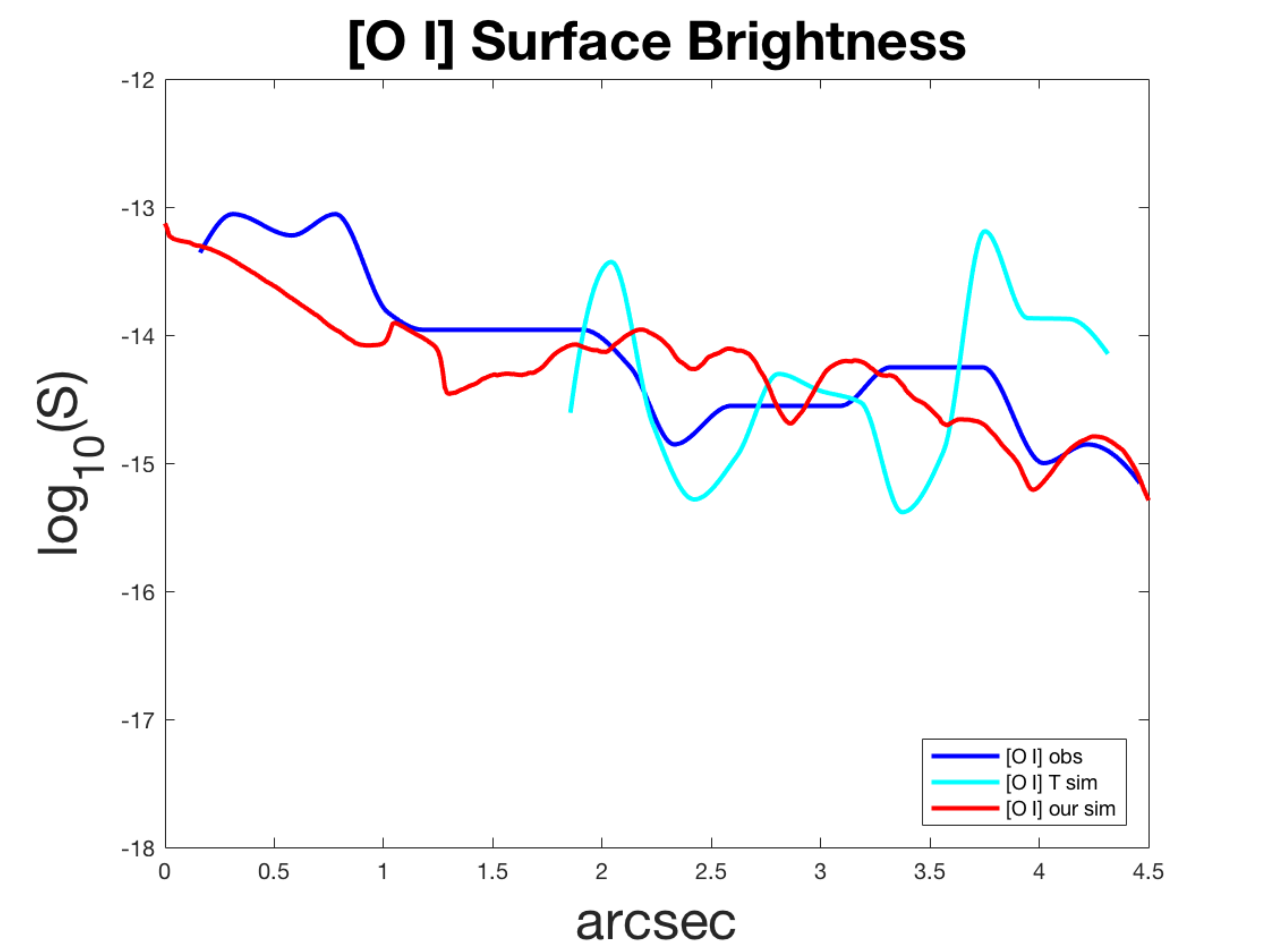}
\caption{Logarithm of surface brightness $\log_{10}(S)$ in the [O~I]
  (6300 + 6363~\AA) doublet along the center of the projected jet,
  with $S$ in erg cm$^{-2}$ arcsec$^{-2}$ s$^{-1}$.  Length scale is
  in arcsec from the source, with 1 arcsec = $0.21 \times 10^{11}$
  km.}
\label{fig:OIcf}
\end{center}
\end{figure}

\begin{figure}
\begin{center}
\includegraphics[scale=0.35]{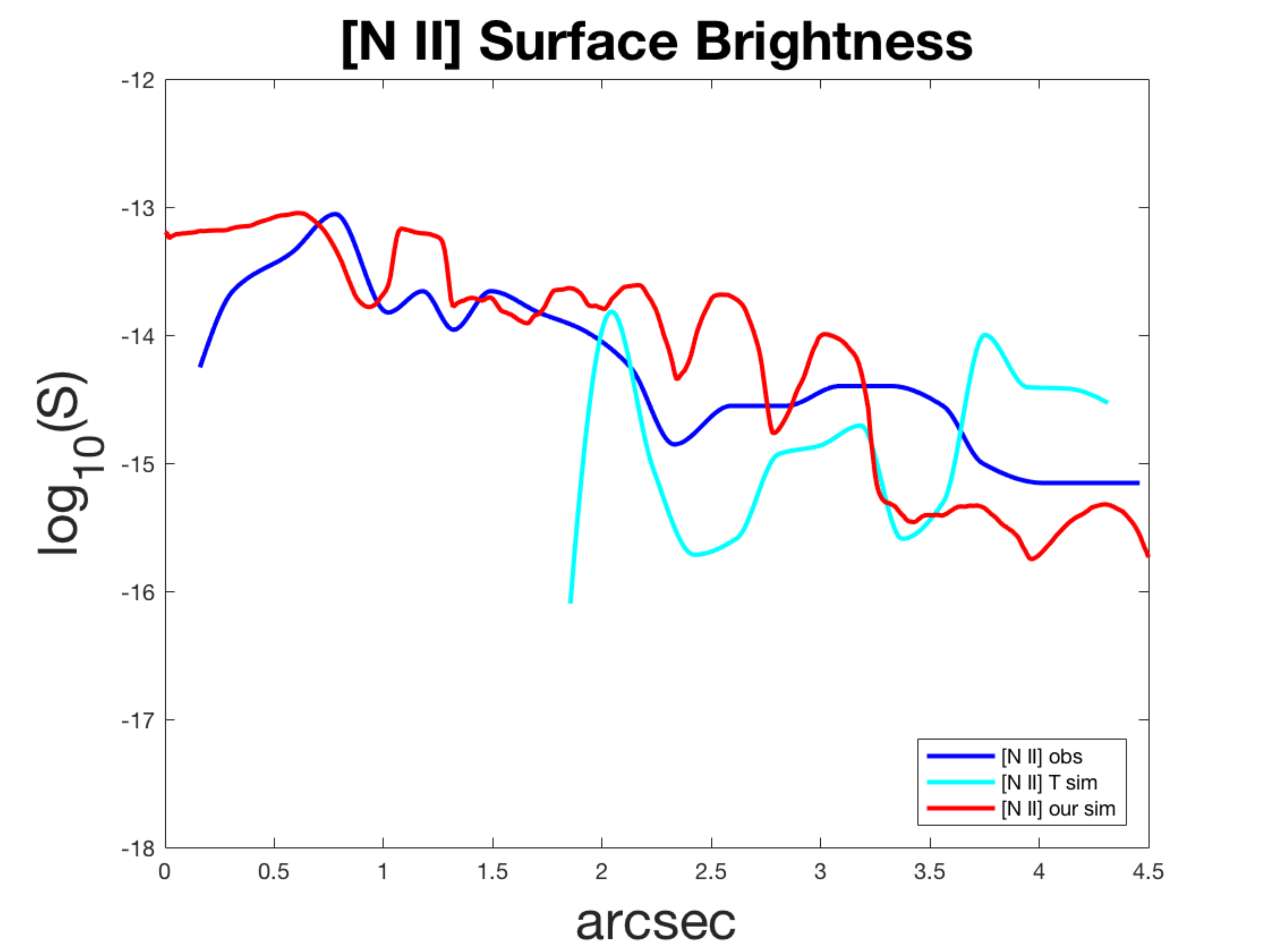}
\caption{Logarithm of surface brightness $\log_{10}(S)$ in the [N~II]
  (6584 + 6548~\AA) doublet along the center of the projected jet,
  with $S$ in erg cm$^{-2}$ arcsec$^{-2}$ s$^{-1}$.  Length scale is
  in arcsec from the source, with 1 arcsec = $0.21 \times 10^{11}$
  km.}
\label{fig:NIIcf}
\end{center}
\end{figure}

\begin{figure}
\begin{center}
\includegraphics[scale=0.35]{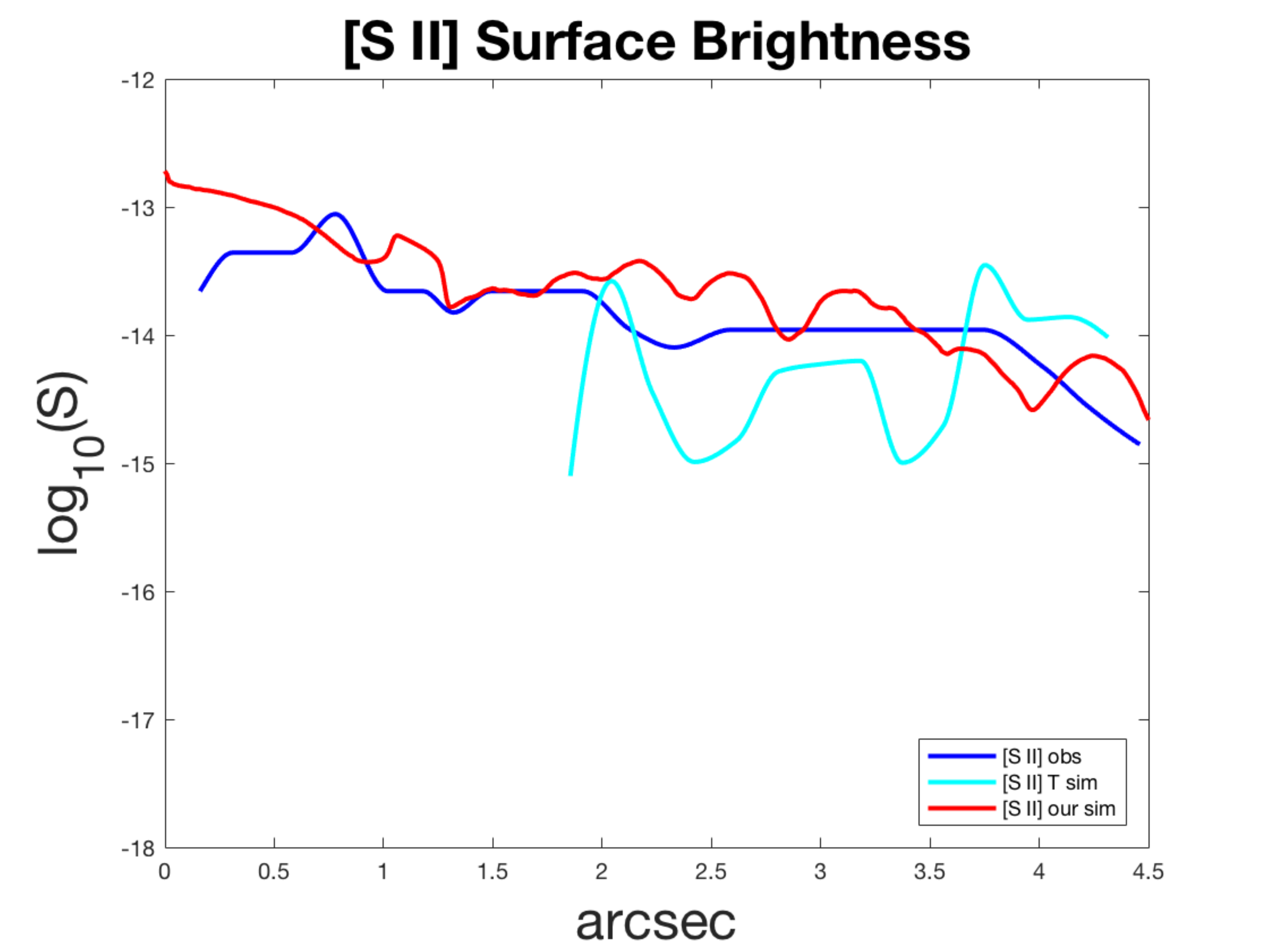}
\caption{Logarithm of surface brightness $\log_{10}(S)$ in the [S~II]
  (6716 + 6731~\AA) doublet along the center of the projected jet,
  with $S$ in erg cm$^{-2}$ arcsec$^{-2}$ s$^{-1}$.  Length scale is
  in arcsec from the source, with 1 arcsec = $0.21 \times 10^{11}$
  km.}
\label{fig:SIIcf}
\end{center}
\end{figure}

Figures~\ref{fig:OIcf}--\ref{fig:SIIcf} present the comparisons of
surface brightness in each of the three doublets along the center of
the projected jet for observational data (labeled ``obs'', from the
data of Hartigan and Morse 2007), MHD multiple-ion simulations
(labeled ``T sim'', Tesileanu et al.\ 2014), and our gas
dynamical/{\em Cloudy}\/ simulations smoothed over 0.2 arcsec (labeled
``our sim'').  Our simulated surface brightness is within $\pm 7.5$\%
on the logarithmic scale of the observational data for the [O~I],
[S~II], and [N~II] forbidden doublets; in addition, the general trend
of our simulated surface brightness maps closely follows the
observational patterns from the limit of the {\em HST}\/ data at 4.5
arcsec all the way up to the circumstellar disk.

\section{Conclusion}

The MHD simulations of Tesileanu et al.\ (2014) calculate radiative
cooling and spectral emission using 19 ionic species: the first three
ionization states (from I to III) of C, N, O, Ne, and S, as well as
H~I, H~II, He~I, and He~II.  Each additional ionic species
significantly increases the computational complexity.  Our atomic
radiative cooling curve uses relevant ionization states of the first
ten most abundant elements in the ISM, and is precomputed and
tabulated: thus the computation of the jet evolution and cooling is
computationally efficient.  Emissivity in any spectral line in {\em
  Cloudy}\/ can be pre-computed as a function of $\log_{10}(n)$ and
$\log_{10}(T)$ and tabulated, and surface brightness in any line
calculated as a post-processing step, which is also computationally
efficient.

There are of course advantages to both approaches, and it is
constructive to compare results of the two approaches.  The approach
of Tesileanu et al.\ (2014) is more fundamental, but our approach,
while more phenomenological, is computationally much faster and can
compute surface brightness in any molecular or atomic line
incorporated into {\em Cloudy}.  Further, simulating astrophysical
jets with a gas dynamical solver with radiative cooling and then
post-processing the simulated densities and temperatures with {\em
  Cloudy}\/ produces simulated surface brightness maps that are in
excellent agreement with the general progression of the observational
data.

\begin{acknowledgements}

We would like to thank Ovidiu Tesileanu for providing the data used in
Figs.~\ref{fig:OIcf}--\ref{fig:SIIcf}, and Evan Scannapieco for
valuable discussions.

\end{acknowledgements}

\section*{References}

\begin{itemize}

\setlength\itemsep{0em}

\item[] Bacciotti, F., Eisl\"offel, J., \& Ray, T.~P. 1999, A\&A 350,
  917

\item[] Ferland, G.~J., Porter, R.~L., van Hoof, P.~A.~M., et
  al.\ 2013, RMxAA, 49, 137

\item[] Gardner, C.~L., \& Dwyer, S.~J. 2009, AcMaS, 29B, 1677

\item[] Gardner, C.~L., Jones, J.~R., \& Hodapp, K.~W. 2016, AJ, 830,
  113

\item[] Gardner, C.~L., Jones, J.~R., Scannapieco, E., \& Windhorst,
  R.~A. 2017, AJ, 835, 232

\item[] Ha, Y., Gardner, C.~L., Gelb, A., \& Shu, C.-W. 2005, JSCom,
  24, 29

\item[] Hartigan, P., \& Morse, J. 2007, AJ, 660, 426 

\item[] Hodapp, K.~W., \& Chini, R. 2014, ApJ, 794, 169

\item[] Hu, X.~Y., Adams, N.~A., \& Shu, C.-W. 2013, JCP, 242, 169

\item[] Kenyon, S., Dobrzycka, D., \& Hartmann, L. 1994, AJ, 108, 1872

\item[] Krist, J.~E., Stapelfeldt, K.~R., Hester, J.~J., et al.\ 2008,
  AJ, 136, 1980

\item[] Le Bourlot, J., Pineau des For\^ets, G., \& Flower,
  D.~R. 1999, MNRAS, 305, 802

\item[] Lopez, R., Riera, A., Raga, et al.\ 1996, MNRAS, 282, 470

\item[] Schmutzler, T., \& Tscharnuter, W.~M. 1993, A\&A, 273, 318

\item[] Shu, C.-W. 1999, High-Order Methods for Computational Physics, Vol.~9
(New York: Springer)

\item[] Stute, M., Gracia, J., Tsinganos, K., Vlahakis, N. 2010, A\&A,
  516, A6

\item[] Tesileanu, O., Matsakos, T., Massaglia, S., et al.\ 2014,
  A\&A, 562, A117

\end{itemize}

\end{document}